\documentclass{ws-procs961x669}
\RequirePackage{graphicx}
\RequirePackage{mathptmx}
\RequirePackage{amsmath}
\RequirePackage{amsfonts}
\RequirePackage{amssymb}
\RequirePackage{latexsym}

\newcommand{\be}{\begin{equation}}
\newcommand{\ee}{\end{equation}}
\newcommand{\bea}{\begin{eqnarray}}
\newcommand{\eea}{\end{eqnarray}}

\renewcommand{\baselinestretch}{1.02}

\begin{document}

\title{Orientability of space from electromagnetic quantum fluctuations}

%\author{N. A. Lemos\thanksref{e1,addr1}
%        \and
 %      \  M. J. Rebou\c{c}as\thanksref{e2,addr2}}

%\thankstext{e1}{e-mail: nivaldolemos@id.uff.br}
%\thankstext{e2}{e-mail: reboucas@cbpf.br}

\author{N. A. Lemos}

\address{Instituto de F\'{\i}sica, Universidade Federal Fluminense,
Av. Litor\^anea, S/N, $\,$24210-340 Niter\'oi -- RJ, Brazil\\
E-mail: nivaldolemos@id.uff.br}

\author{M. J. Rebou\c cas}

\address{Centro Brasileiro de Pesquisas F\'{\i}sicas,
Rua Dr.\ Xavier Sigaud 150,
$\,$22290-180 Rio de Janeiro -- RJ, Brazil\\
E-mail: reboucas.marcelo@gmail.com}

\date{Received: date / Accepted: date}

%\maketitle
%\sloppy

\begin{abstract}
Whether the $3-$space where we live is a globally orientable manifold $M_3$, and
whether the local laws of physics require that $M_3$  be equipped with
a canonical  orientation, are among the important unsettled questions in
cosmology and quantum field theory. % and in cosmology.  fundamental physics
It is often assumed that a test for spatial orientability requires a global
journey across the whole $3-$space to check for orientation-reversing
closed paths. Since such a global expedition is not feasible, physically
motivated theoretical arguments are usually offered to support the choice
of canonical time orientation for the  $4-$dimensional spacetime manifold,
and space orientation for $3-$space.
One can certainly take advantage of such theoretical arguments to
support these  assumptions on orientability, %of spacetime manifolds, %%canonical
but the ultimate answer should rely on cosmological observations or local
experiments, or can come from a topological fundamental theory of physics.
In a recent paper we have argued that it is potentially possible to
locally access the the $3-$space orientability of Minkowski empty
spacetime through physical effects involving point-like 'charged'
objects under vacuum quantum electromagnetic fluctuations.
More specifically, we have studied the stochastic motions of a
charged particle and an electric dipole subjected to these fluctuations
in Minkowski spacetime, with either an orientable or a non-orientable
$3-$space topology, and derived analytical expressions for a statistical
orientability indicator in these two flat topologically inequivalent manifolds. %topologies
For the charged particle, we have shown that it is  possible to distinguish
the two topologies by contrasting the evolution of their respective indicators.
For the point electric dipole we have found that a characteristic inversion
pattern exhibited by the curves of the orientability indicator is a signature
of non-orientability, making it  possible to locally probe
the orientability of Minkowski $3-$space in itself. % (no need of comparison).
Here, to shed some additional  light on the spatial orientability, we briefly
review these results, and also discuss some of its features and consequences.
The  results might be seen as opening the way to a conceivable experiment involving
quantum vacuum electromagnetic fluctuations to look into the spatial
orientability of Minkowski empty spacetime.
\end{abstract}


%\PACS{03.70.+k \and 05.40.Jc \and 42.50.Lc \and 04.20.Gz \and 98.80.Jk \and 98.80.Cq}

%\PACS{04.50.Kd, 98.80.-k, 95.36.+x}

\bodymatter

%%%%%%%%%%%%%%%%%%%%%%%%%%%%%%%%%%%%%%%%%%%%%%%%%%%%
\section{Introduction}\label{Intro}
%%%%%%%%%%%%%%%%%%%%%%%%%%%%%%%%%%%%%%%%%%%%%%%%%%%%

In the framework of general relativity the Universe is described as a
four-dimensional differentiable manifold  locally endowed %
with a spatially homogeneous and isotropic Friedmann--Lema\^{\i}tre--%
Robertson--Walker (FLRW) metric.
Geometry is a local attribute that brings about intrinsic curvature, whereas
topology is a global feature of a manifold related, for example, to compactness
and orientability.
The FLRW spatial geometry constrains but does not specify the topology
of the spatial sections, $M_3$, of the space-time manifold, which  we
assume to be of the form $\mathcal{M}_4 = \mathbb{R} \times M_3$.
In the FLRW description of the Universe, two diverse sets of fundamental
questions are related to, first, the $3-$geometry; second, to
the topology of spatial sections $M_3$.
Regarding the latter, at the cosmological level it is expected that one should be able
to detect the spatial topology through cosmic microwave background radiation (CMB)
or (and) stochastic primordial gravitational waves~\cite{CosmTopReviews,Reboucas2020}.
%which should follow some basic detectability conditions~\cite{TopDetec}.
However, so far, direct searches for a nontrivial topology of $M_3$, using
CMB data from Wilkinson Microwave Anisotropy Probe (WMAP) and
Planck collaborations, have found no convincing evidence of
nontrivial topology below the radius of the last scattering surface%
~\cite{Planck-2013-XXVI,Planck-2015-XVIII,Cornish-etal-04,Key-etal-07, Bielewiz-Banday-11,Vaudrevange-etal-12,Aurich-Lustig-13}.
This absence of evidences  does not exclude the possibility of a FLRW~ %\refcite
universe with a detectable nontrivial spatial topology~\cite{Bernui-etal-18,
Aurich_etal-21,Gomero-Mota-Reboucas-2016}.

It is well-known that the topological properties of a manifold antecede
its geometrical features and the differential tensor structure with which
the physical theories are formulated.
Thus, it is relevant to determine whether, how  % formulated.
and to what extent physical results depend upon or are somehow affected by
a nontrivial topology.
Since the net role played by the spatial topology is more properly examined
in  static space-times,  the dynamical degrees of freedom of which are frozen,
here we focus on the static Minkowski space-time, whose spatial % in this work
geometry is Euclidean.
However,  rather than the general topology of the spatial sections $M_3$ of
Minkowski space-time, in this work we investigate its topological property
called \textsl{orientability}, which is related to the existence of
orientation-reversing closed paths on the spatial section $M_3$.
Questions as to whether the $3-$space of Minkowski space-time, which is the standard
arena of quantum field theory,  is necessarily an orientable manifold, or to what
extent the known \textsl{laws of physics require a canonical spatial orientation} are
among the underlying primary concerns of this work.             %us to equip it

To be more precise regarding the setting of the present  work, let us first
briefly provide some mathematical results. % background. details.
The spatial section $M_3$ of the Minkowski spacetime  manifold
$\mathcal{M}_4 = \mathbb{R}\times M_3$ is usually taken to be the %assumed
simply-connected Euclidean space $\mathbb{E}^{3}$, but it is known mathematical
result
that it can also be any one of the possible $17$ topologically distinct
quotient manifolds $M_3 = \mathbb{E}^3/\Gamma$, where $\Gamma$ is a
discrete group of isometries or holonomies acting freely on the covering
space $\mathbb{E}^{3}$.\footnote{
For recent accounts on the classification of three-dimensional
Euclidean spaces the reader is referred to
Refs.~\refcite{Wolf67}~--\refcite{Fujii-Yoshii-2011}.}
The action of $\Gamma$ tiles the covering manifold $\mathbb{E}^{3}$ into
identical cells which are copies of what is known as fundamental domain (FD)
fundamental cell or polyhedron (FC or FP).
So, the multiple-connectedness gives rise to periodic conditions (repeated cells)
in the simply-connected covering manifold $\mathbb{E}^{3}$ that are defined
by the action  of the group $\Gamma$ on $\mathbb{E}^{3}$.
Different groups $\Gamma$ give rise to different periodicities on
the covering manifold $\mathbb{E}^{3}$, which in turn define different
Euclidean spatial topologies $M_3$ for Minkowski spacetime.
These mathematical results make it explicit that besides the
simply-connected $\mathbb{E}^{3}$ there is a variety of
topological possibilities ($17$ classes) for the spatial section
of Minkowski spacetime. % $M_3$
The potential consequences of multiple-connectedness for physics
come about when, for example,  % possible, potential
one takes  into account that in a manifold with periodic boundary conditions
only certain modes of fields can exist. In this way, a nontrivial topology
may leave its signature on the expectation values of local physical
quantities\cite{Bessa-Reboucas-2020}.
So, for example, the energy density for a scalar field in Minkowski
space-time with nontrivial spatial section is shifted from  the corresponding
result for the Minkowski space-time with trivial spatial topology.
This is the  Casimir effect of topological origin%
~\cite{dhi79,Dowker-Critchley-1976,st06,MD-2007,DHO-2002,DHO-2001}.

Regarding orientability, the central issue of this work,
three important points should be emphasized.
First, we mention that it is widely assumed, implicitly or explicitly,
that a four-dimensional manifold $\mathcal{M}_4 = \mathbb{R}\times M_3$
that models the physical world is spacetime orientable and, additionally,
that it is separately time and space orientable.
Second, eight out of the above-mentioned %mathematically classified
$17$ quotient flat $3-$manifolds are non-orientable%
~\cite{Wolf67,Thurston,Adams-Shapiro01,Cipra02,Riazuelo-et-el03,Fujii-Yoshii-2011}.
A non-orientable $3-$space is then a concrete mathematical possibility
among the quotient manifolds $M_3 = \mathbb{E}^3/\Gamma$, and it comes about
when the  holonomy group $\Gamma$ contains at least a flip or reflexion
as one of its elements. % of the holonomy group $\Gamma$.
Finally, it is generally assumed that, being a global property, the $3-$space
orientability cannot be tested locally. In this way, to disclose the spatial
orientability of our physical world one would have to make trips along some
specific closed paths  around  $3-$space to check, for example,
whether one returns with left- and right-hand sides exchanged.

Since such a global journey across the whole $3-$space is not feasible one
might think that spatial orientability cannot be probed.
In this way, one would have either to answer the orientability question
through cosmological observations or local experiments.
Hence, assuming that  spatial orientability is a falsifiable
property of   $3-$space, a question that arises is whether
spatial orientability can be subjected to local tests.
%% %subjected to local tests.%%%experimental
%%
Our main goal in this work is to present a way to tackle  this question.
To this end,  we have investigated\cite{Lemos-Reboucas2021} stochastic
motions of a charge particle and
an electric dipole under vacuum quantum fluctuations of the electromagnetic field
in Minkowski spacetime with two inequivalent spatial topologies, namely
the non-orientable \textsl{slab with flip} and the orientable \textsl{slab},
which are often denoted by the symbols $E_{17}$ and  $E_{16}$, respectively%
~\cite{Riazuelo-et-el03,Fujii-Yoshii-2011}.
Manifolds endowed with these topologies turned out to be   % manifolds
appropriate to identify orientability or non-orientability signatures through
the stochastic motions of point-like particles in Minkowski
spacetime.
In the next section, to shed some additional light on spatial orientability,
we briefly review the most significant results of our article,
Ref.~\refcite{Lemos-Reboucas2021}, and  discuss some of their consequences.

%%%%%%%%%%%%%%%%%%%%%%%%%%%%%%%%%%%%%%%%%%%%%%%%%%%%%%%
\section{Main results and their interpretation}  \label{TopSet}
%%%%%%%%%%%%%%%%%%%%%%%%%%%%%%%%%%%%%%%%%%%%%%%%%%%%%%%%

The general idea underlying our work is to perform a comparative study
of the time evolution of  physical systems in Minkowski spacetime with
orientable and non-orientable spatial sections. To this end, the physical
systems chosen are a point charged particle and  a point electric dipole
under vacuum quantum electromagnetic fluctuations. We shall describe the
results for the dipole only because they are more significant
--- see Ref.~\refcite{Lemos-Reboucas2021} for the point charge case.

%%%%%%%%%%%%%%%%%%%%%%%%%%%%%%%%%%%%%%%%%%%%%%%%%%%%
\subsection{The setting and the physical system}
%%%%%%%%%%%%%%%%%%%%%%%%%%%%%%%%%%%%%%%%%%%%%%%%%%%%

We begin by recalling that simply-connected spacetime manifolds
are necessarily orientable. On the other hand, the product of two manifolds is
simply-connected if and only if the factors are.
Thus, the space-orientability of  Minkowski spacetime manifold
$\mathcal{M}_4 = \mathbb{R}\times M_3$ reduces to  orientability
of the $3-$space $M_3$.
In this paper, we shall consider the topologically nontrivial spaces $E_{16}$ and $E_{17}$.
The slab space $E_{16}$  is constructed by tessellating $\mathbb{E}^3$ by equidistant
parallel planes, so it has only one compact dimension associated with a  direction
perpendicular to those planes. Taking the $x$-direction as compact,  one has that,
with $n_x\in \mathbb{Z}$ and $a>0$, points $(x,y,z)$ and  $(x+n_xa,y,z)$ are identified
in the case of the slab space $E_{16}$.
The slab space with flip $E_{17}$ involves an additional inversion of a direction orthogonal
to the compact direction, that is, one direction in the tessellating planes is flipped
as one moves from one plane to the next. Letting the flip be in the $y$-direction,
 the identification of points  $(x,y,z)$ and $(x+n_xa,(-1)^{n_x}y,z)$ defines the
 $E_{17}$ topology.
The slab space  $E_{16}$ is  orientable whereas the slab space with
flip $E_{17}$ is non-orientable.

The physical system we consider consists of a point electric dipole
with  moment $\bf p$ and mass $m$ that is locally subject to %left to move under
vacuum fluctuations of the electric field ${\bf E}({\bf x}, t)$
in Minkowski spacetime $\mathcal{M}_4$
with the metric $\eta_{\mu\nu}=\mbox{diag} (+1, -1, -1, -1)$.
The topology of the spatial section $M_3$ is taken to be
either $E_{16}$ (orientable) or $E_{17}$ (non-orientable) instead
of  $\mathbb{E}^{3}$.

%We  consider a  nonrelativistic point electric dipole with
% mass $m$ and  electric dipole moment $\bf p$ which is locally subject to
%vacuum fluctuations of the electric field ${\bf E}({\bf x}, t)$
%in a topologically nontrivial spacetime manifold
%%$\mathcal{M}_4$, which is decomposed into $\mathcal{M}_4 = \mathbb{R}\times M_3$
%equipped with the Minkowski metric $\eta_{\mu\nu}=\mbox{diag} (+1, -1, -1, -1)$.
%% Colocar em algum ponto: In this paper we use units $\hbar = c =1$.
%The spatial section is usually taken to be $\mathbb{E}^{3}$,
%but here we take for $M_{3}$ either $E_{16}$ or $E_{17}$.

The nonrelativistic motion of the  dipole is locally determined by
%%the equation of motion
\begin{equation}
\label{eqmotion-dipole}
m\frac{d{\bf v}}{dt} = {\bf p}\cdot {\mbox{\boldmath $\nabla$}}{\bf E}({\bf x},t)\,,
\end{equation}
%%%%
where  $\mathbf{v}$ is the dipole's velocity and  %dipole's
$\mathbf{x}$ its position at  time $t$.
We assume that the dipole practically does not move on the time scales of interest.
Thus the dipole has a negligible displacement, and  we  can ignore the time
dependence of  $\mathbf{x}$. So, $\mathbf{x}$ is taken to be constant in what
follows~\cite{yf04,lrs16}.%

%We assume that on the time scales of interest the dipole practically does not
%move, that is, it has a negligible displacement, so we  can ignore the time
%dependence of  $\mathbf{x}$.
%Thus, the  dipole's position $\mathbf{x}$ is taken as constant in what
%follows~\cite{yf04,lrs16}.%
Assuming the dipole is initially
at rest,  integration of Eq.~\eqref{eqmotion-dipole} yields
%%%%
\begin{equation}\label{eqmotion-dipole-integrated}
{\bf v}({\bf x}, t) = \frac{1}{m}p_j\int_0^{t}\partial_j{\bf E}({\bf x}, t^{\prime})\,dt^{\prime}
\end{equation}
with $\partial_j = \partial/\partial x_j$ and summation over repeated indices implied.
%%%%

The mean squared speed in each of the three independent directions
$i = x, y, z$ is given by
\begin{equation}\label{eqdispersion-dipole}
\Bigl \langle\Delta v^2_i\Bigr \rangle = \frac{p_jp_k}{m^2} \int_0^t\int_0^t
\Bigl \langle \bigl(\partial_j E_i({\bf x}, t')\bigr)
\bigl(\partial_k E_i({\bf x}, t'')\bigr) \Bigr \rangle\, dt' dt''\,,
\end{equation}
which can be conveniently rewritten as
\begin{equation}\label{eqdispersion-dipole-rewritten}
\Bigl \langle\Delta v^2_i\Bigr \rangle = \lim_{{\bf x}^{\prime} \to {\bf x}}\frac{p_jp_k}{m^2}
\int_0^t\int_0^t \partial_j\partial_k^{\,\prime} \Bigl \langle  E_i({\bf x}, t')
 E_i({\bf x}', t'')\Bigr \rangle\, dt' dt''
\end{equation}
where $\partial_k^{\,\prime}= \partial/\partial x_k^{\prime}$ and there
is no summation over $i$.
Following Ref.~\refcite{yf04}, we assume that the electric field is %% Yu and Ford~\cite{yf04
a sum of quantum  $\mathbf{E}_q$ and classical  $\mathbf{E}_c$  parts.
Since ${\bf E}_c$ is not subject to quantum fluctuations and
$\langle {\bf E}_q\rangle =0$,
the two-point function $\langle E_i({\bf x}, t)E_i({\bf x}', t')\rangle$ in
equation~\eqref{eqdispersion-dipole-rewritten} involves only
the quantum part $\mathbf{E}_q$ of the electric field. %~\cite{yf04}.
% $\mathbf{E}= \mathbf{E}_c + \mathbf{E}_q \,$.

In Minkowski spacetime with a topologically nontrivial spatial section,
the {spatial separation} $r^2$ that enters the electric field correlation
functions takes a  form that captures the
periodic boundary conditions imposed on the covering space $\mathbb{E}^{3}$
by the covering group  $\Gamma$, which characterize the spatial topology.
In consonance with Ref.~\refcite{st06}, the spatial separations for $E_{16}$
and $E_{17}$  are
\begin{eqnarray}
\label{separation-E16}
E_{16}: \qquad r^2 & = & (x - x'- n_x a)^2 + (y - y')^2 + (z - z')^2, \\
\label{separation-E17}
E_{17}: \qquad r^2 & = & (x-x'-n_x a)^2+(y -(-1)^{n_x} y')^2
+ (z - z')^{2}.
\end{eqnarray}

%%%%%%%%%%%%%%%%%%%%%%%%%%%%%%%%%%%%%
\subsection{Orientability indicator}
%%%%%%%%%%%%%%%%%%%%%%%%%%%%%%%%%%%%%

The orientability indicator  that we will consider is defined by replacing
the electric field correlation functions in Eq.~(\ref{eqdispersion-dipole-rewritten})
by their renormalized counterparts\cite{Lemos-Reboucas2021}.
For a dipole oriented along the $y$-axis the dipole moment is ${\bf p}=(0,p,0)$ and
we have
\begin{equation}\label{eqdispersion-dipole-y}
 ^{^{(y)}}{\!} \mbox{\large $I$}_{v^2_i}^{E_{17}}({\bf x},t) =
  \lim_{{\bf x}^{\prime} \to {\bf x}}\frac{p^2}{m^2}
\int_0^t\int_0^t \partial_y\partial_{y^{\prime}} \Bigl \langle  E_i({\bf x}, t')
 E_i({\bf x}', t'')\Bigr \rangle_{ren}^{E_{17}}\, dt' dt'' \,,
\end{equation}
where the left superscript within parentheses  indicates the dipole's orientation.

The renormalized correlation functions are given by\cite{Lemos-Reboucas2021}
\begin{equation}\label{correlation-i-E17}
\bigl \langle E_i({\bf x}, t)E_i({\bf x}', t')\bigr \rangle_{ren}  =
\frac{\partial }{\partial x_i} \frac{\partial}
{\partial {x'}_i}D_{ren}({\bf x}, t; {\bf x}', t')  - \frac{\partial }{\partial t} \frac{\partial}
{\partial t'}D_{ren} ({\bf x }, t; {\bf x'}, t')
\end{equation}
where
\begin{equation}\label{Hadamard-ren}
D_{ren}({\bf x}, t; {\bf x}', t') =
\sum\limits_{{n_x=-\infty}}^{{\infty\;\;\prime}}\frac{1}{4\pi^2(\Delta t^2 - r^2)}
\end{equation}
and where $\Delta t = t-t^{\prime}$,  $\,\sum_{}^{\;'}$ indicates that
the Minkowski contribution term $n_x = 0$ is excluded from the summation,
and the spatial separation for $E_{17}$ is given by Eq.~\eqref{separation-E17}.

%%in which $\sum_{}^{\;'}$ indicates that
%%the Minkowski contribution term $n_x = 0$ is excluded from the summation,
%%$\Delta t = t-t^{\prime}$, and
%the spatial separation for $E_{17}$ is given by Eq. \eqref{separation-E17}.
The Hadamard function $D({\bf x}, t; {\bf x}', t') $ for the multiply-connected
space is defined by including the term with $n_x = 0$ in the
summation~\eqref{Hadamard-ren}. Thus, $D_{ren} = D - D_0$, where $D_0$
is the Hadamard function for the simply-connected  space.

Before calculating the components of the orientability
indicator in equation~\eqref{eqdispersion-dipole-y},
%%Before presenting the results for some components of the orientability
%%indicator~\eqref{eqdispersion-dipole-y},
we point out that from equations~\eqref{eqdispersion-dipole-rewritten}
and \eqref{eqdispersion-dipole-y}~--~\eqref{Hadamard-ren} one can figure out
a general definition of the orientability indicator as
\begin{equation}  \label{new-ind}     %% \mathlarger\Delta  \,
 \mbox{\large $I$}_{v^2_i}^{MC}
= \Bigl\langle\Delta v_i^2 \Bigr \rangle^{MC}
- \;\,\Bigl \langle\Delta v_i^2 \Bigr \rangle^{SC} ,
\end{equation}
where  $\bigl\langle\Delta v_i^2 \bigr \rangle$  is the
mean square velocity dispersion, and the superscripts $SC$ and $MC$ %and
stand for simply- and multiply-connected manifold, respectively.
The right-hand side of~(\ref{new-ind}) is defined by first taking the difference
of the two terms with ${\bf x}^{\prime} \neq  {\bf x}$ and then setting
${\bf x}^{\prime} =  {\bf x}$ (coincidence limit).

The components of the orientability indicator for the dipole in $E_{17}$
are~\cite{Lemos-Reboucas2021}
\begin{equation}\label{eqdispersion-dipole-yx-final-E17}
 ^{^{(y)}}{\!} \mbox{\large $I$}_{v^2_x}^{E_{17}}({\bf x},t)= -\frac{4p^2}{\pi^2m^2} \sum\limits_{{n_x=-\infty}}^{{\infty\;\;\prime}}(-1)^{n_x}\bigg\{ 2I_1
 + 3 (r^2 - r_x^2 + 6r_y^2)I_2 + 24 (r^2 - r_x^2) r_y^2I_3   \bigg\},
\end{equation}
\begin{eqnarray}\label{eqdispersion-dipole-yy-final-E17}
 ^{^{(y)}}{\!} \mbox{\large $I$}_{v^2_y}^{E_{17}}({\bf x},t) &  = & -\frac{2p^2}{\pi^2m^2}
\sum\limits_{{n_x=-\infty}}^{{\infty\;\;\prime}} (-1)^{n_x}
\bigg\{ (5-3(-1)^{n_x}) I_1 \nonumber \\
& + &   6 [r^2 + (7-6(-1)^{n_x})r_y^2] I_2 + 48 [r^2 -(-1)^{n_x}r_y^2]r_y^2I_3 \bigg\}
\end{eqnarray}
and
\begin{equation}\label{eqdispersion-dipole-yz-final-E17}
 ^{^{(y)}}{\!} \mbox{\large $I$}_{v^2_z}^{E_{17}}({\bf x},t)  =  -\frac{4p^2}{\pi^2m^2} \sum\limits_{{n_x=-\infty}}^{{\infty\;\;\prime}}(-1)^{n_x}\bigg\{ 2I_1  + 3(r^2 + 6 r_y^2 )I_2
 + 24 r^2 r_y^2I_3 \bigg\}.
\end{equation}
where, with $\Delta t = t' - t''$,
\begin{equation}\label{integral-1}
 I_1 =  \int_0^t  \int_0^t \frac{dt^{\prime}dt^{\prime\prime}}{(\Delta t^2 -r^2)^3}
= \frac{t}{16}\bigg[ \frac{4t}{r^4 (t^2-r^2)} +\frac{3}{r^5} \ln \frac{(r-t)^2}{(r+t)^2} \bigg],
\end{equation}
\begin{equation}\label{integral-2}
 I_2 =  \int_0^t  \int_0^t \frac{dt^{\prime}dt^{\prime\prime}}{(\Delta t^2 -r^2)^4}
= \frac{1}{6r}\frac{\partial I_1}{\partial r} = \frac{t}{96}
\bigg[ \frac{4t(9r^2-7t^2)}{r^6 (t^2-r^2)^2}  - \frac{15}{r^7} \ln \frac{(r-t)^2}{(r+t)^2} \bigg],
\end{equation}
\begin{equation}\label{integral-3}
 I_3 =  \int_0^t  \int_0^t \frac{dt^{\prime}dt^{\prime\prime}}{(\Delta t^2 -r^2)^5}
= \frac{1}{8r}\frac{\partial I_2}{\partial r}   = \frac{t}{768}\bigg[
 \frac{4t(57t^4- 136r^2t^2 + 87r^4)}
{r^8 (t^2-r^2)^3} + \frac{105}{r^9} \ln \frac{(r-t)^2}{(r+t)^2} \bigg].
\end{equation}
In Eqs.~(\ref{eqdispersion-dipole-yx-final-E17}) to
(\ref{integral-3}) one must put
\begin{eqnarray}
\label{r-rx-ry-coincidence-E17}
 r & = & \sqrt{n_x^2a^2 + 2(1-(-1)^{n_x})y^2}, \nonumber \\
 r_x^2 & = & n_x^2a^2,
\qquad   r_y^2 = 2(1-(-1)^{n_x})y^2.
\end{eqnarray}

The components of the dipole orientability  indicator
for the slab space $E_{16}$
are obtained from those for $E_{17}$  by setting $r_x^2 = r^2, r_y=0 $,
and replacing $(-1)^{n_x}$ by $1$ everywhere.
Therefore, we have
\begin{eqnarray}\label{eqdispersion-dipole-yx-final-E16}
^{^{(y)}}{\!} \mbox{\large $I$}_{v^2_x}^{E_{16}}({\bf x},t) & = & -\frac{8p^2}{\pi^2m^2}
\sum\limits_{{n_x=-\infty}}^{{\infty\;\;\prime}} I_1 ,\\
\label{eqdispersion-dipole-yy-final-E16}
^{^{(y)}}{\!} \mbox{\large $I$}_{v^2_y}^{E_{16}}({\bf x},t) & = & -\frac{4p^2}{\pi^2m^2}
\sum\limits_{{n_x=-\infty}}^{{\infty\;\;\prime}} ( I_1 + 3r^2I_2 ),\\
\label{eqdispersion-dipole-yz-final-E16}
^{^{(y)}}{\!} \mbox{\large $I$}_{v^2_z}^{E_{16}}({\bf x},t) & = & -\frac{4p^2}{\pi^2m^2}
\sum\limits_{{n_x=-\infty}}^{{\infty\;\;\prime}} (2I_1 + 3r^2I_2 ),
\end{eqnarray}
in which $r=\vert n_x \vert a$.

%%
%%%%%%%%%%%%%%%%%%%%%%%%%%%%%%% Fig.1  Dipole  %%%%%%%%%%%%%%%%%%%%%%%%%%%%%%%%%
\begin{figure*}[t!]          %% {E16_vs_E17-fig4} %%
\begin{center} %\centering
\includegraphics[width=7.1cm,height=5.9cm]{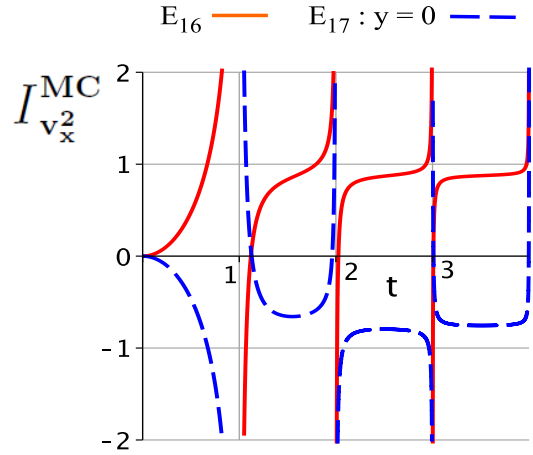}
%{dipole_fig1_MC_New2.eps} %%% [scale=0.4]
 \hspace{8mm}  %
%%%%           %%%%            %%%%             %%%%         %%%%
\begin{minipage}[t]{\textwidth} \vspace{-1mm}
\renewcommand{\baselinestretch}{0.96}
\caption{Time evolution of the  orientability
indicator \eqref{eqdispersion-dipole-yx-final-E17},  in units of $p^2/m^2$,
for a  point electric dipole with mass $m$ and dipole moment $p$,  oriented in the flip $y-$direction, in Minkowski spacetime with
non-orientable $E_{17}$ and orientable $E_{16}$ spatial topologies, with 
compact length $a=1$.
The solid and dashed lines stand for the indicator  curves  %dispersion
for a dipole in $3-$space with $E_{16}$ and $E_{17}$ topologies, respectively.
For the inhomogeneous topology $E_{17}$ the dipole is %placed
at $P_0= (x,0,z)$ to freeze out the topological  inhomogeneity degree of freedom\cite{Lemos-Reboucas2021}.
Both orientability indicator  curves do present a periodicity pattern, 
but the curve for the non-orientable $E_{17}$ exhibits a different kind of 
periodicity characterized by  a distinctive inversion pattern.
Non-orientability is responsible for this  pattern of sequential inversions,
which is absent in the indicator curve for the orientable $E_{16}$.   %dispersion
\label{E16_vs_E17-fig4}  }  \vspace{2mm}
\end{minipage}
\end{center}
\end{figure*}

%%%%%%%%%%%%%%%%%%%%%%%%%%%%%%%%%%%%%%%%%%%%%%%%%%%%%%%%%%%%%%
\subsection{Analysis of the results}\label{Dipole-conclusive}
%%%%%%%%%%%%%%%%%%%%%%%%%%%%%%%%%%%%%%%%%%%%%%%%%%%%%%%%%%%%%%

\begin{figure*}[tb]              %% (E16_vs_E17-fig5}
\begin{center} %\centering
\includegraphics[width=7.1cm,height=5.9cm]{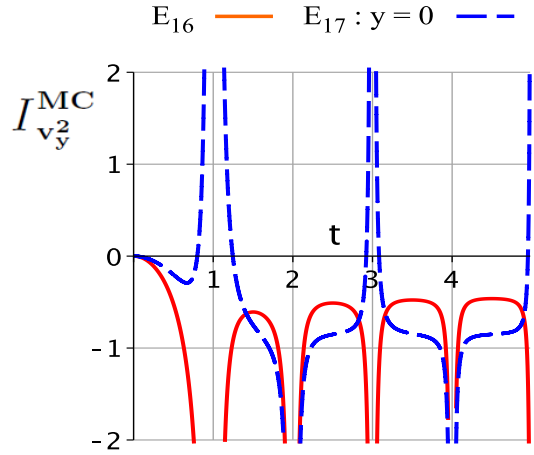} %%% [scale=0.4]
 \hspace{8mm}  %
%\includegraphics[width=7.1cm,height=5.9cm]{dipole_fig4.eps}  %%  [scale=0.4]
%%%%           %%%%            %%%%             %%%%         %%%%
\begin{minipage}[t]{\textwidth} \vspace{-1mm}
\renewcommand{\baselinestretch}{0.96}
\caption{Time evolution of the orientability indicator \eqref{eqdispersion-dipole-yy-final-E17} under the same
conditions as those  of Fig.~\ref{E16_vs_E17-fig4}.
The orientability indicator curve for
$E_{17}$ again exhibits a characteristic inversion pattern but now 
different from the one for the $v_x$-component shown in Fig.~\ref{E16_vs_E17-fig4}.
For the $v_y$-component displayed  here the non-orientability signature  can be recognized in the 
pattern of consecutive upward and 
downward horn-like figures formed by the dashed  curve.
\label{E16_vs_E17-fig5}  }  %\vspace{5mm}
\end{minipage}
\end{center}
\end{figure*}

Since equations  (\ref{eqdispersion-dipole-yx-final-E17})%
-(\ref{eqdispersion-dipole-yz-final-E17})  and (\ref{eqdispersion-dipole-yx-final-E16})-(\ref{eqdispersion-dipole-yz-final-E16})
are too complicated to allow a straightforward interpretation, we plot
figures for the components of the orientability indicator.
Figures~\ref{E16_vs_E17-fig4} and \ref{E16_vs_E17-fig5} arise from
Eqs.~~\eqref{eqdispersion-dipole-yx-final-E17}~--~\eqref{eqdispersion-dipole-yz-final-E17}
as well as \eqref{eqdispersion-dipole-yx-final-E16}~--~\eqref{eqdispersion-dipole-yz-final-E16},
with the topological length $a=1$ and  $n_x \neq 0$ ranging from $-50$ to $50$.
In Fig.~\ref{E16_vs_E17-fig4} and Fig.~\ref{E16_vs_E17-fig5}  the solid lines stand
for the curves of the orientability indicator for the dipole in  Minkowski spacetime
with $E_{16}$ orientable spatial topology,
whereas the dashed lines correspond to the curves of the orientability indicator curves
for the dipole located at $P_0 = (x,0,z)$ in a $3-$space with $E_{17}$ non-orientable
topology.

In the case of the $x$-component, the time evolution curves of the orientability indicator
for $E_{16}$ and $E_{17}$, shown in Fig.~\ref{E16_vs_E17-fig4}, present a common periodicity
but with  distinguishable patterns. The orientability indicator curve for $E_{17}$ displays
a distinctive sort of periodicity characterized by  an \textsl{inversion pattern}.
Non-orientability gives rise to  this  pattern of successive inversions, which does not
occurs in the case of the orientable $E_{16}$.

The differences become more conspicuous  when one considers  the $y$-component of %striking
the orientability indicator, shown in Fig.~\ref{E16_vs_E17-fig5}.
The non-orientability of $E_{17}$ is disclosed by an
inversion pattern whose structure is more striking than the one for the $x$-component.
The orientability indicator curves for $E_{17}$ form a  pattern of alternating upward
and downward  % dispersion
``horns'',  making  the non-orientability of $E_{17}$ unmistakably identifiable.
The $z$-component of the indicator
 behaves the same way~\cite{Lemos-Reboucas2021}.

The characteristic inversion pattern exhibited exhibited by the dipole
indicator curves makes it possible to identify the non-orientability of the
manifold $E_{17}$ in itself, with no need of a comparison
%% without having to make a
with the indicator curves for its orientable counterpart $E_{16}$.
However, this sort of 
comparison is necessary in the point charge case, as 
discussed in detail in  Ref.~\refcite{Lemos-Reboucas2021}.

In brief,  it may be possible to  unveil %%the above discussion suggests  that
a putative  non-orientability of the spatial section of Minkowski spacetime
by local means, namely by the stochastic motions of
charged point-like particles caused by  electromagnetic quantum vacuum fluctuations.
If the motion of a point electric dipole is taken as probe, non-orientability
can be intrinsically detected by the inversion pattern of the dipole curves
of the orientability indicator \eqref{new-ind}.

 %%% AQUI

%%%%%%%%%%%%%%%%%%%%%%%%%%%%%%%%%%% Fig.2 Dipole %%%%%%%%%%%%%%%%%%%%%%%%%%%
%%%%%%%%%%%%%%%%%%%%%%%%%%%%%%%%%%%%%%%%%%%%%%%%%%%%%%%%%%%%%%%%%%%%%%%%%%%%%%

%%%%%%%%%%%%%%%%%%%%%%%%%%%%%%%%%%%%%%%%%%%%%%%%%%%%%%%%%%
\section{Concluding remarks}  \label{Conclusion}
%%%%%%%%%%%%%%%%%%%%%%%%%%%%%%%%%%%%%%%%%%%%%%%%%%%%%%%%%%%

In general relativity and quantum field theory  spacetime is modeled as
a differentiable manifold, which is a topological space endowed with a
differential structure.  
It is generally assumed that the spacetime manifolds involved in these
frameworks are entirely orientable, meaning  that they are separately space and time  orientable. 
The theoretical arguments used to adopt these assumptions on orientability
combine the space-and-time universality of the basic local rules of physics
with physically well-defined (thermodynamically) local arrow of time,
CP violation and CPT invariance%
~\cite{Zeldovich1967,Hawking,Geroch-Horowitz-1979}.
One can certainly use such reasonings in support of the standard assumptions
on the global orientability structure of spacetimes.%
\footnote{See Ref. \refcite{MarkHadley-2018} and related Ref.~\refcite{MarkHadley-2002} 
for a different point of view on this matter.}
Nevertheless, it is legitimate to expect that the ultimate answer to
questions regarding the orientability of spacetime manifolds should rely 
on astro-cosmological observations or local experiments, or even might 
come from  a fundamental topological theory in physics.

In the physics at daily and even astrophysical length and time scales, 
we do not encounter signs or hints of non-orientability. This being true,
the  open question that remains is whether the physically well-defined local 
orientations can be extended to microscopic or cosmological scales. 

At the cosmological scale, one would think at first sight that to disclose  spatial
orientability one would have to make a trip around the whole $3-$space to check for
orientation-reversing paths.
However, a  determination of the spatial topology (detection and 
subsequent reconstruction of the topology)  through the so-called 
circles in the sky~\cite{CSS1998}, for example, would bring out as a bonus 
an answer to the $3-$space orientability problem at the cosmological scale.
However, no convincing evidence of nontrivial spatial topology below the radius of 
the last scattering surface has been found up to now\cite{Planck-2013-XXVI,Planck-2015-XVIII,Cornish-etal-04,Key-etal-07, Bielewiz-Banday-11,Vaudrevange-etal-12,Aurich-Lustig-13}.

Parallel to these works, in this article we have addressed the question as to whether 
electromagnetic quantum vacuum  fluctuations can be used to bring out  the 
spatial orientability of Minkowski spacetime, in which the possible effects 
of the dynamical degree of  freedom of FLRW spacetime is  frozen in order 
to separate the topological and dynamical roles on the orientability problem. %features.  
We have
found that there exists a characteristic inversion pattern exhibited by the
the curves of our  orientability indicator \eqref{new-ind} for a dipole in the case of
$E_{17}$, signaling that the non-orientability of $E_{17}$ can be detected by
 per se, that is, with no need for comparisons. 
Thus, the inversion pattern of the  orientability indicator curves
for the dipole is a signature of  the reflection holonomy, which is expected to be
present in the orientability indicator curves for the dipole in all remaining seven
non-orientable topologies that contain a flip holonomy, namely  the four Klein 
spaces ($E_{7}$ to $E_{10}$) and  the chimney-with-flip class ($E_{13}$ 
to $E_{15}$).

Observation of physical phenomena and controlled experiments are essential 
to our understanding of nature.
Our results indicate the possibility of a local experiment to unveil
spatial non-orientability through  stochastic motions of point-like particles %objects
under electromagnetic quantum vacuum fluctuations.   %%quantum vacuum 
The present paper may be seen as a hint of a possible
way to locally probe the spatial orientability of Minkowski  spacetime.

%%%%%%%%%%%%%%%%%%%%%%%%%%%
%%%%%%%%%%%%%%%%%%%%%%%%%%%
\section*{Acknowledgements}
%%%%%%%%%%%%%%%%%%%%%%%%%%
M.J. Rebou\c{c}as acknowledges the support of FAPERJ under a
\textsl{CNE E-26/202.864/2017} grant, and thanks CNPq for the
grant \textsl{306104/2017-2} under which this work was carried out.
We also thank  A.F.F. Teixeira and C.H.G.  Bessa for fruitful
discussions.
%%%%%%%%%%%%%%%%%%%%%%%%%%%

%%%%%%%%%%%%%%%%%%%%%%%%%%%%%%

%%%%%%%%%%%%%%%%%%%%%%

\end{document}